# Efficient perturbations for basin hopping in amorphous glasses

Coraline Du,[1,2] Hye Sol Kim,[2] Scott C. Warren[2]

[1]*Cuthbertson High School, Waxhaw, NC 28173, USA,*

[2]*Department of Chemistry, Kenan Laboratory, UNC Chapel Hill, Chapel Hill, NC 27599, USA*

Efficient exploration of the complex potential-energy landscapes of amorphous materials is central to computational structure discovery and refinement. Conventional Monte Carlo, reverse Monte Carlo, and related methods typically sample configuration space through small, local trial moves and may require millions to tens of millions of moves to converge. Here, we evaluate larger, nonlocal perturbations followed by local geometry relaxation as an alternative sampling strategy. We develop and test four perturbation types using amorphous $Al_2O_3$ as a model system. Among them, moving an oxygen atom to change the coordination numbers of two aluminum atoms, allows access to low-energy configurations with substantially fewer trial moves than a conventional Monte Carlo trajectory. These results suggest that relaxation-assisted nonlocal moves could reduce trapping in local minima and improve sampling in structure-search and reverse Monte Carlo workflows.

## I. INTRODUCTION

Amorphous materials can exhibit properties distinct from those of their crystalline counterparts, yet connecting these properties to atomic structure remains difficult because the absence of long-range periodicity limits the direct application of conventional crystallographic structure-solution methods. Total-scattering measurements retain information about short- and intermediate-range order, but reconstructing a unique three-dimensional atomic model from a radially averaged scattering signal is intrinsically underdetermined. Reverse Monte Carlo (RMC) addresses this inverse problem by iteratively perturbing atomic coordinates and accepting or rejecting trial configurations according to their agreement with experimental scattering data.[1] Because structurally different configurations can reproduce similar pair correlations, however, data-driven RMC can produce models containing unrealistic coordination environments, bond-angle distributions, or network topologies.[2] Hybrid reverse Monte Carlo (HRMC) reduces this ambiguity by incorporating an interatomic-energy term alongside the experimental misfit.[3] In our recent work, we introduced pyHRMC, a Python package that applies HRMC to electron total-scattering data, and demonstrated that it recovered a target amorphous $Al_2O_3$ structure more faithfully than conventional RMC.[4]

Although energetic and chemical constraints can improve the physical realism of refined structures, the outcome of a Monte Carlo search also depends strongly on the trial moves used to explore configuration space. Common RMC and HRMC implementations rely primarily on Metropolis-style local proposals, such as a small random displacement of one atom followed by acceptance or rejection according to the resulting objective function.[1,3,5] These moves are straightforward and permit incremental refinement, but they alter the structure only locally. The potential-energy landscape of a glass contains many local minima organized into basins, and transitions between topologically distinct configurations may require coordinated rearrangements that are unlikely to arise from a sequence of independent, small displacements.[6] Other structure-search methods illustrate the value of more disruptive proposals. Wooten–Winer–Weaire bond-switch moves directly reorganize the topology of amorphous networks,[7] whereas basin-hopping and minima-hopping methods combine larger structural changes with local energy minimization to move among inherent structures.[8,9] These approaches motivate the use of nonlocal, relaxation-assisted trial moves in experiment-constrained refinement, where the design of the proposal distribution may be as consequential as the form of the objective function.

Here, we isolate the role of trial-move design by comparing **N** distinct structural perturbations, each followed by local geometry relaxation, using amorphous $Al_2O_3$ as a model system. Amorphous alumina provides a demanding test case because its disordered network accommodates multiple Al coordination environments and a broad distribution of local geometries.[10] We evaluate how effectively each perturbation lowers the relaxed energy and enables transitions among energy basins relative to a conventional sequence of small atomic displacements. Among the perturbations examined, moving an oxygen atom to change the coordination numbers around two aluminum atoms allows the system to reach low-energy configurations with substantially fewer trial moves than the conventional Monte Carlo sequence. These results identify trial-move design as an independent means of improving configurational exploration and provide a basis for incorporating nonlocal, relaxation-assisted proposals into pyHRMC and related methods that combine experimental scattering constraints with energetic information.[4]

## II. METHODS

### A. Movement types

We developed four perturbation algorithms to modify an initial sample of amorphous alumina. The applications used are: pymatgen.core, Crystal NN, The algorithms are as follows: Coordination Balance Perturber, Random Swap Perturber, Random Angle Perturber, and Shared Polyhedra Perturber.

*Coordination Balance Perturber:* This perturbation aims to redistribute oxygen atoms to change the structure's topology. The algorithm selects an oxygen atom near a highly coordinated aluminum atom and moves it to a less coordinated aluminum atom.

*Random Angle Perturber:* This perturbation slightly adjusts bond angles between two oxygen atoms bonded to an aluminum atom. The algorithm randomly selects an aluminum atom and finds two oxygen atoms between 1.2 and 2.1 Å . The aluminum atom is the center of rotation, and one oxygen atom is rotated between -13° and +13°. The other oxygen is fixed. This perturbation results in a change in the bond angle between an O-Al-O triplet.

*Random Atom Swap:* The algorithm randomly selects an aluminum atom and an oxygen atom and swaps the coordinates of those atoms without modifying the coordinates of any other atoms.

*Shared Polyhedra Perturber:* This perturbation changes the local connectivity between two aluminum atoms. The algorithm finds the oxygen atoms bonded to the aluminum atoms and randomly generates a pair of aluminum atoms that "share" oxygen atoms. To increase sharing, an oxygen atom that is already bonded to an aluminum is displaced into the region between the Al pair so they are bonded to both atoms. To decrease sharing, an oxygen atom that is already bonded to two aluminums is displaced away from the sharing region. The perturbation is only accepted if it achieves the intended shared oxygen value, a minimum bond distance, and if no atoms overlap. This perturbation is controlled by the six submodes: *corner-to-edge, corner-to-face, edge-to-corner, edge-to-face, face-to-corner,* and *face-to-edge*.

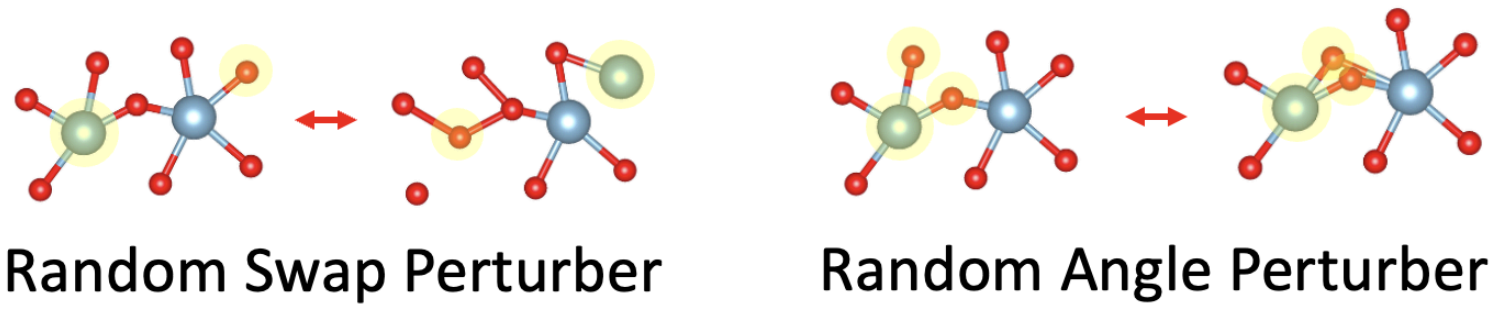

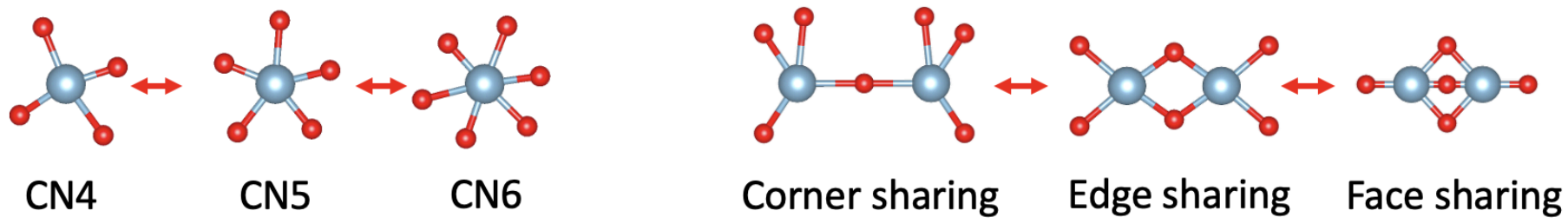

Figure 1. Structural perturbation modes used to explore the configurational landscape of Al2O3. Red atoms represent oxygen, and blue atoms represent aluminum.

### B. Relaxation methods

Structural relaxations were performed using a neuroevolution potential (NEP) [11] through the *calorine* `CPUNEP` calculator. Atomic positions and cell parameters were simultaneously optimized under periodic boundary conditions using the FIRE algorithm and ASE's `FrechetCellFilter`.

Relaxation was continued at zero external pressure until the maximum force was below 0.03 eV/Å, with a maximum of 2,000 optimization steps.

### C. Test sequences

Unconditional acceptance algorithm: In the first testing approach, every perturbation was accepted regardless of whether it increased the energy. This unconditional approach was used to evaluate the volatility of a movement and how the structure would evolve without any constraints. A single method was used to sequentially perturb the structure one hundred times. After loading the initial structure and assigning its parent energy, a perturbation to the structure (*child)* would be attempted. If the perturbation failed more than ten iterations, it would be skipped. Following the perturbation, the child structure would be immediately relaxed using NEP. The energy, delta energy from the previous perturbation, coordination number and its amount, perturbation method, and metadata containing the specific atoms perturbed were recorded. Next, the child structure would be promoted to be the new parent structure and continue the iterative loop.

Energy-based acceptance algorithm: In the second approach, the same algorithm was used; however, with a constraint. After relaxing the *child* structure, the energy of the structure was compared to the parent energy. If the energy was greater than the parent energy, the perturbation was rejected. This approach allows us to see how quickly a movement can lower the energy and for how long the method remains effective.

### D. Starting geometries

High-energy amorphous $Al_2O_3$ structures were generated using molecular dynamics (MD) driven by a neuroevolution potential (NEP). Starting from amorphous $Al_2O_3$ configurations, elevated-temperature (3500 K) MD was used to sample structurally distorted configurations spanning a broad range of potential energies. Selected configurations were subsequently geometry optimized using the same NEP potential. Both atomic positions and simulation-cell degrees of freedom were relaxed using the FIRE optimizer with a FrechetCellFilter until the maximum atomic force was below 0.03 eV $Å^{-1}$, with a maximum of 2000 optimization steps. The resulting relaxed structure represents distinct local minima on the NEP potential-energy landscape and were used as high-energy starting configurations for subsequent structural sampling/refinement. The initial structure had a relaxed energy of -760.32 eV. Because the initial structure contained no face-sharing polyhedra, face-to-edge and face-to-corner perturbations were not performed.

## III. RESULTS AND DISCUSSION

### Unconditional testing:

In this test, we explored how the energy changes when a specific movement type occurs (Figure 2, Table 1). We performed the movement type repeatedly on the same supercell, thus allowing us to compare how efficiently different movement types approach a low energy configuration. By comparing replicate runs of the same type, we examined the reproducibility of these findings.

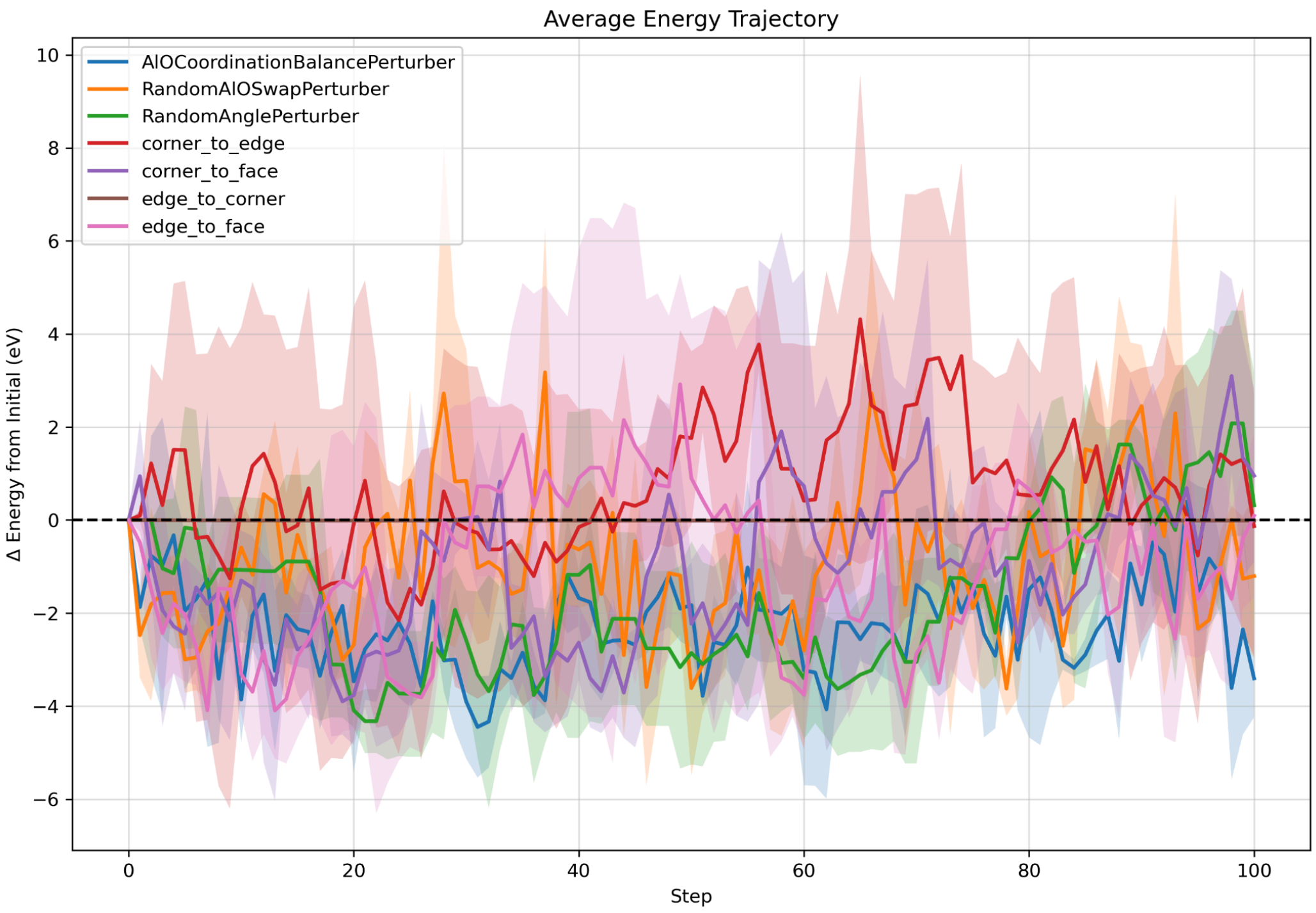

Figure 2. Average energy trajectory for a single movement type. The lines indicate the average of three different trials with random seeds. The shading shows the three individual runs. The coordination balance perturber pictured in blue finds the lowest energy state and has the lowest energy at the end of testing. Corner-to-edge increases the energy the most; it reaches the highest peak and has the highest minima.

| **Movement type** | **Mean $\Delta E$ from previous state (eV)** | **Standard deviation of $\Delta E$ (eV)** |
|---|---:|---:|
| Coordination balance | −0.0340 | 1.7123 |
| Random angle | 0.0032 | 1.1928 |
| Edge-to-face | 0.0009 | 1.6161 |
| Corner-to-face | 0.0095 | 1.7690 |
| Random swap | −0.0120 | 2.4902 |
| Edge-to-corner | −0.0001 | 0.0006 |
| Corner-to-edge | −0.0013 | 1.5777 |

Table 1. Statistics from averaging the three individual runs of a movement type. Mean and standard deviation of the energy change between consecutive relaxed structures ($\Delta E_previous$) averaged over all perturbation steps and three independent trials.

Negative values indicate a decrease in energy, while positive values indicate an increase in energy. The overall trend in the data shows that, on average, most movements change the energy of the structure by a recordable amount. The coordination balance perturber, on average, reduced the energy by -0.034 eV every move. These results suggest this is the result of the large topology change that is created when altering the coordination number of two aluminum atoms. The movement redistributes oxygen, which allows for a big drop in energy after relaxation. The initial perturbations can allow the structure to overcome an energy basin and drop to a lower state. The random swap perturber also performed well, with a delta energy of -0.012 eV. One possible explanation is the incredibly disruptive movement of swapping atoms. The edge-to-face and corner-to-edge perturbations on average changed the energy by a small amount. Edge to corner had the least impact on the structure's energy (-0.0001 eV).

The standard deviations describe the volatility of the movement type. The coordination balance perturber has moderate to high variability (1.71 eV), meaning that it is both effective and somewhat reliable. In contrast, random swap perturbation exhibited the most variability (2.49 eV), indicating that it explored a wide range of energies, but on average lowered energy. This is due to the very disruptive nature of the movement, swapping atoms can either change the structure to a very favorable configuration or a very unfavorable configuration. The edge-to-corner perturber had a very low standard deviation, meaning that it has very low variability and performs almost identically every run.

These statistics allow us to draw three conclusions. One, the coordination balance perturber is the most effective movement to lower energy each step and is fairly consistent. An explanation could be due to the highly disruptive nature of the movement, which allows the structure to overcome the energy basin and reach new minima. Due to the limited pairs of aluminum the perturber can choose, it reduces the variability of the movement. Second, the edge-to-corner movement, on average, is the least effective movement to change energy, because of the smallest absolute value of delta previous. Third, random swap had the most volatility by far, making it a very positive or negative movement.

**Energy-based testing:**
The results of this test allow us to identify the most efficient movement for lowering a structure's energy with constraints. Since movements that only reduce energy are accepted, we can observe which movements reach the lowest state, and how many steps it takes until the perturbation plateaus.

| Movement type | Mean $\Delta E$ from previous state (eV) | Standard deviation of $\Delta E$ (eV) |
|---|---|---|
| Coordination balance | −0.6694 | 1.2730 |
| Random angle | −0.1511 | 0.5458 |
| Corner-to-face | −0.2816 | 0.6311 |
| Edge-to-face | −0.2177 | 0.5196 |
| Corner-to-edge | −0.1984 | 0.4404 |
| Random swap | −0.5995 | 0.7887 |
| Edge-to-corner | −0.0005 | 0.0005 |

Table 2. Global comparison of energy change for different movement types.

Based on the data, on average, the most efficient move is the coordination balance perturber (Table 2). On average, it lowers the energy by -0.67 eV per step and has high variability, meaning it can lower the energy even more. However, the movement that finds the lowest energy state is the random angle perturber. At the end of 3 trials, the movement lowered energy by around 7 eV. While the coordination balance perturber is the most efficient, it is not the most effective. This is because, over the three trials, the coordination balance perturber stopped running after 5-10 steps, lowering the energy drastically in very few steps. The random angle perturber lowered less per step, but ran for 35-50 steps. The random angle perturber takes considerably longer to reach the lowest energy state. Because some delta energy is close to zero, one might initially interpret that relaxation is lowering the energy instead of perturbations. However, looking at the individual runs, there is quite a lot of variability. Barring the outlier edge-to-corner, there is a 4 eV range between the lowest energy and highest energy produced by the movements. And since there is a moderate amount of standard deviation, the perturbation is still a factor. Therefore, while relaxation minimizes the energy after a step, the perturbation itself is the main factor in how low the energy exploration can reach.

The coordination balance perturber causes a massive shift in the overall topology of the structure. The movement of one oxygen across different coordinated regions causes the oxygen to bond differently. From initial testing, amorphous aluminum oxide prefers to bond in tetrahedral and octahedral shapes. The increase in connectivity between Al-O correlates to a more stable structure. So the movement of coordination number can first increase the energy of the structure and then, followed by relaxation, find a new lower-energy state of the structure. Because the coordination balance ran ten or fewer moves, this movement type is only effective for huge jumps over a few perturbations. The results suggest that the perturbation either runs out of viable local regions to perturb, or simply finds the lowest state it can with the movement. The latter is likely true since there are forty aluminum atoms in the structure; there is not a scarcity of available pairs. There is an abundance of coordination number 5 at the beginning of the run; after the run, the amount of coordination number 4 increased, which is a more stable shape for amorphous alumina. However, because the coordination balance only redistributes a higher-coordinated oxygen to a lower-coordinated oxygen, the number of octahedral Al-O in the structure can only decrease, which can limit the energy exploration of the movement type.

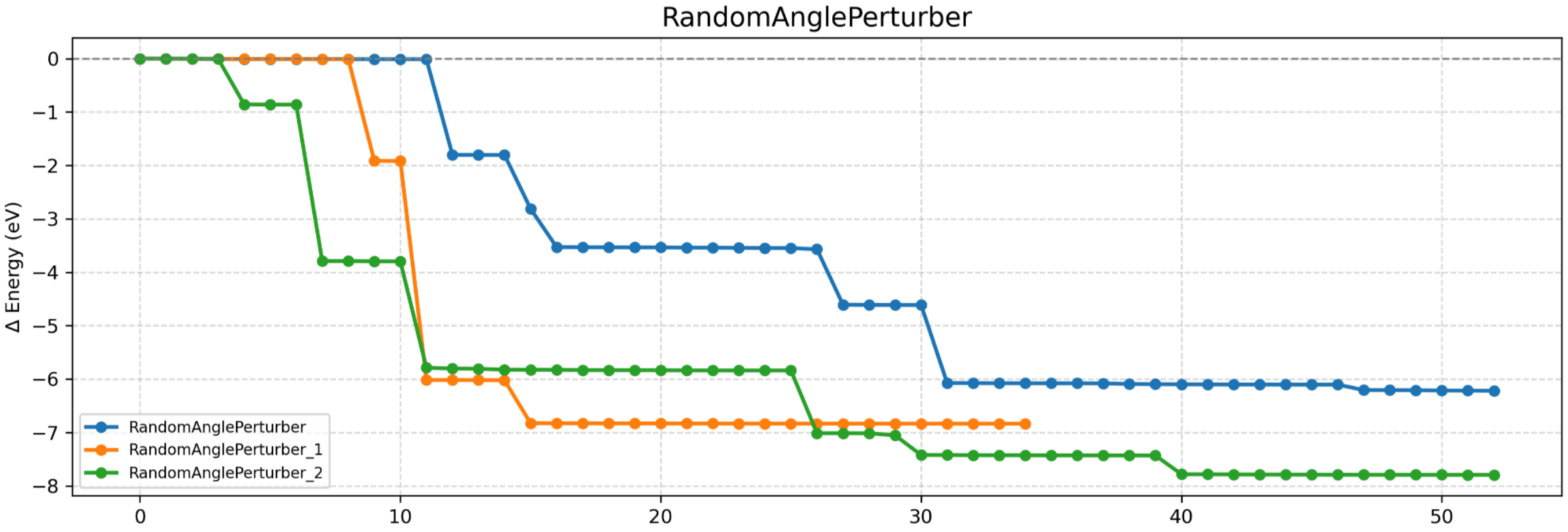

Figure 3. Three runs consisting solely of random angle perturbations.

The random angle perturber causes minor shifts in the structure, which allows the trial to run for longer, allowing for more chances for the energy to be lowered (Figure 3). The movement is smaller compared to others; it only perturbs one oxygen atom from -13 to +13 degrees. After relaxation, the atoms find the optimal position and lower the energy slightly. In Figure 3, we can observe that the perturbations reach a plateau, and then there is a sudden energy shift. The relaxation is likely returning the perturbation to a very similar energy stage until a random pair is perturbed, which allows the structure topology to change drastically. However, because of relaxation, this should return the perturbation to the local minima. The lowest energy reached during all of the testing is pictured in green, with an energy of -768.11 eV. Because of the energy plateaus, this movement is incredibly inefficient. The shared polyhedra perturbers all executed somewhat similarly. In Table 2, the corner-to-face, edge-to-face, and corner-to-edge movements had a delta eV of -0.28, -0.21, and -0.20, respectively. We can observe from the results that shared polyhedra perturbers that increase sharing between two atoms lower the energy of the structure. The greater increase in sharing correlates to a larger energy drop. While edge-to-corner, which decreased sharing, almost kept the energy the same. Looking at the edge-to-corner plot, there was a consistent energy that the perturbation would end at: -760.28, which is the highest energy found during testing. This small (-0.000495 eV) change was consistent over three trials (standard deviation of 0.00048 eV); each trial ranged from 21 to 25 total steps. The results from the shared polyhedra perturber and coordination balance perturber suggest that in amorphous aluminum oxide, movements that increased polyhedra connectivity are associated with a distinct energy drop in the structure. A local movement (changing the sharing) and a non-local movement (redistributing coordination) cause amorphous structures to have a more interconnected Al-O topology, which can lead to a more stable structure. The edge-to-corner perturber is the only movement purposely reducing sharing and polyhedra connectivity, therefore leading to a very minuscule decrease in energy.

## IV. CONCLUSIONS

This study demonstrates that the design of structural perturbations strongly influences how efficiently relaxation-assisted searches explore the potential-energy landscape of amorphous Al2O3. Among the perturbations examined, the coordination balance perturber was the most efficient at lowering the energy. By moving an oxygen atom between aluminum coordination environments, it produced large changes in network topology and substantial energy reductions within ten or fewer accepted moves. The random angle perturber, in contrast, was the most effective at locating the lowest-energy configuration observed in this study. Its smaller, more local structural changes permitted a greater number of accepted perturbations and ultimately reached an energy of −768.11 eV, although it required many more steps and frequently produced extended energy plateaus. Thus, the coordination balance and random angle perturbations offer complementary advantages: the former rapidly reaches a lower-energy region of configuration space, whereas the latter explores that region more gradually and can locate still-lower-energy minima.

The results also reveal a relationship between Al–O network connectivity and the relaxed energy. Perturbations that increased oxygen sharing between aluminum-centered polyhedra generally produced larger energy decreases, with greater increases in sharing corresponding to larger reductions in energy. In comparison, the edge-to-corner perturbation, which reduced polyhedral sharing, produced almost no net energy change and repeatedly returned the structure to nearly the same relaxed energy. Together with the performance of the coordination balance perturber, these observations indicate that perturbations that reorganize the structure toward a more interconnected Al–O topology provide a particularly effective means of accessing lower-energy configurations.

These complementary perturbations could be combined in future Monte Carlo or reverse Monte Carlo workflows. A coordination balance perturbation could first be used to produce rapid, large-scale changes in topology, particularly in larger structures where many distinct coordination environments are available. Subsequent random angle perturbations could then provide finer exploration within the lower-energy region reached by the initial nonlocal moves. Future work should test mixed-movement sequences directly and determine whether alternating local and nonlocal perturbations improves sampling relative to trajectories based on a single movement type. It would also be useful to modify the coordination balance algorithm so that it explicitly favors four- and six-coordinate aluminum environments while reducing the population of five-coordinate aluminum, and then evaluate how this constraint affects both energy exploration and the resulting network structure. Because several trajectories reached extended energy plateaus, future implementations should additionally investigate acceptance criteria that require a meaningful energy decrease or otherwise distinguish genuine transitions between minima from perturbations that relax back to nearly identical configurations. These developments would provide a basis for incorporating efficient, relaxation-assisted nonlocal moves into experiment-constrained Monte Carlo refinement methods.